\documentstyle[preprint,aps,epsf,floats]{revtex}
\def\xslash#1{{\rlap{$#1$}/}}
\def\dsl{\,\raise.15ex\hbox{/}\mkern-13.5mu D}
\def\lqcd{\Lambda_{\rm QCD}}
\def\OMIT#1{{}}
\def\finite#1{+\dots}
\def\finiteomit#1{{}}

\tighten
\begin{document}

\preprint{ \vbox{\hbox{UTPT-00-03} \hbox{hep-ph/0005275} } }

\title{Summing Sudakov logarithms in $B\rightarrow X_s \gamma$ in
  effective field theory} \author{Christian W. Bauer\footnote{{\tt
      bauer@physics.utoronto.ca}}, Sean Fleming\footnote{{\tt
      fleming@physics.utoronto.ca}}, Michael Luke\footnote{{\tt
      luke@physics.utoronto.ca}}}
\address{Department of Physics, University of Toronto \\
  60 St.~George St., Toronto, Ontario, Canada M5S1A7}
\date{May 2000}

\maketitle
\begin{abstract}
  We construct an effective field theory valid for processes in which
  highly energetic light-like particles interact with collinear and
  soft degrees of freedom, using the decay $B\rightarrow X_s\gamma$
  near the endpoint of the photon spectrum, $x = 2 E_\gamma / m_b \to
  1$, as an example.  Below the scale $\mu=m_b$ both soft and
  collinear degrees of freedom are included in the effective theory,
  while below the scale $\mu=m_b\sqrt{x-y}$, where $1-y$ is the
  lightcone momentum fraction of the $b$ quark in the $B$ meson, we
  match onto a theory of bilocal operators.  We show that at one loop
  large logarithms cancel in the matching conditions, and that we
  recover the well known renormalization group equations that sum
  leading Sudakov logarithms.
\end{abstract}

\pagebreak

\section{Introduction}

Effective field theories (EFT's) provide a simple and elegant method
for calculating processes with several relevant energy
scales\cite{eftpapers}.  Part of the utility of EFT's is that 
they dramatically simplify the summation of powers of logarithms of
ratios of mass scales, which would otherwise make perturbation theory
poorly behaved.  For example, in a theory with a very heavy particle
of mass $M$, one-loop corrections will typically be enhanced by 
$\log(M/\lambda)$, where $\lambda$ is a low scale in the problem.  In
the EFT in which the heavy particle has been removed from the theory,
such logarithms are replaced by factors of $\log(\mu/\lambda)$ (where
$\mu$ is the renormalization scale in dimensional regularization, or
the cutoff in cutoff regularization), and the complete series of
leading logarithms $\alpha_s^n \log^n(\mu/\lambda)$ is straightforward to
sum via the renormalization group.

The situation is more complicated for processes with highly energetic
light particles.  In this case, there are both collinear and infrared
divergences in the theory, which give rise to the familiar Sudakov
double logarithms \cite{sudafed}.  For example, the perturbative
expansion of the $N$'th moment of the photon spectrum in inclusive
$b\rightarrow X_s \gamma$ decay is of the form
\begin{equation}
\sum_n \sum_{m\leq 2n} C_{n,m}\alpha_s^n\log^m N.
\end{equation}
Although the arguments of these logarithms are not obviously the ratio of
two scales, they arise because the typical energy and invariant
mass of light particles are widely separated, and they may be summed
via well-known techniques based on factorization theorems~\cite{css1}
into the form 
\begin{equation}\label{nseries}
\exp \left[ \sum _n \left( a_n \alpha_s^n \log^{n+1} \! N +  b_n
    \alpha_s^n \log^{n} \! N \right) + \dots \right] \,.
\end{equation}
The terms $\alpha_s^n \log^{n+1} \! N$ are referred to as the leading
logarithmic contribution, the terms $\alpha_s^n \log^{n} \! N$ are
referred to as the next-to-leading logarithmic contribution, and the
remaining terms are called subdominant. 

Recently there has been some discussion in the literature of summing
Sudakov logarithms using effective field theory
techniques~\cite{gk,aglietti1,bmk}.  Such an approach could have
several advantages over the conventional method; in particular, while
factorization formulas are based on perturbation theory, EFTs,
by construction, are valid beyond perturbation theory, and by including
higher dimension operators it
should be straightforward (if tedious) to go beyond
the leading twist approximation.  In the various versions of the EFT
approach which have been suggested, the effective theory is the
so-called ``Large Energy Effective Theory'' (LEET)~\cite{dg}, which
describes light-like particles coupled to soft degrees of freedom.
However, a difficulty with the approaches presented to date is that,
as pointed out in Refs.~\cite{bmk}, in the minimal subtraction (MS)
scheme logarithms arising at one loop in LEET do not match logarithms
arising at one loop in QCD for any choice of the matching scale $\mu$;
hence these logarithms may not be summed using the RGE's.

In this paper we consider this problem in the context of $B\rightarrow
X_s \gamma$ decays.\footnote{In fact, the authors of \cite{llr} argued
that the resummation of subleading Sudakov logarithms is not
necessary for practical purposes for this decay.  Nevertheless, it
provides a simple example in which we may compare our results 
to those in the literature.}
We show that the problem of matching scales may
be resolved by introducing a new intermediate effective theory
containing both soft and collinear degrees of freedom, which is then
matched onto LEET (effectively integrating out the collinear modes) at
a lower scale.  We show that the matching conditions onto both
effective theories contain no large logarithms at one loop.  We then
calculate the RGE's in the two theories summing the leading logarithms
and a certain subset of the next-to-leading logarithms. To this order the
expression obtained for the resummed Sudakov logarithms is identical
to that derived in Refs~\cite{ks,ar}.

\section{Sudakov Logarithms in $B\rightarrow X_S\gamma$ and LEET}
\label{scales}

Inclusive decays of heavy quarks have been well understood for many
years in the context of an operator product expansion (OPE) in the
inverse mass of the heavy quark\cite{hqetope}.  At leading order in
the $\lqcd/m_b$ expansion the $B$ meson decay rate is equal to the $b$
quark decay rate, and nonperturbative effects are suppressed by at
least two powers of $\lqcd / m_b$. However, the OPE only converges for
sufficiently inclusive observables. Unfortunately, experimental cuts on
measurements of rare decays such as $B \to X_s \gamma$, $B \to X_s
\ell^+ \ell^-$, and $B\rightarrow X_u\ell\bar\nu$ are required,
restricting the available phase space considerably. Since all of these
decays are of phenomenological interest, either in the determination of
$|V_{ub}|$ or detection of new physics, understanding inclusive decays
in restricted regions of phase space is important. 

If the phase space
is restricted such that the final hadronic state is dominated by only
a few resonances, the breakdown of the OPE simply reflects the fact that an inclusive
treatment based on local duality is no longer appropriate.  This is
the case for the dilepton invariant mass spectrum in inclusive $B \to
X_s \ell^+ \ell^-$ and $B \to X_u \ell \bar\nu$
decays~\cite{bll}. However, when the kinematic cut is in a region of
phase space dominated by highly energetic, low invariant mass final
states, the OPE breaks down even for quantities smeared over a
parametrically larger region of phase space, where the decay is not
resonance dominated.  This situation arises in the endpoint region of
the electron energy spectrum and the low hadronic invariant mass
region in semileptonic $B\rightarrow X_u\ell\bar\nu$ decay, as well as
the endpoint region of the photon spectrum in $B\rightarrow X_s
\gamma$ decay\cite{llr,endpoint}.

\begin{figure}[htbp]
  \epsfxsize=14cm \hfil\epsfbox{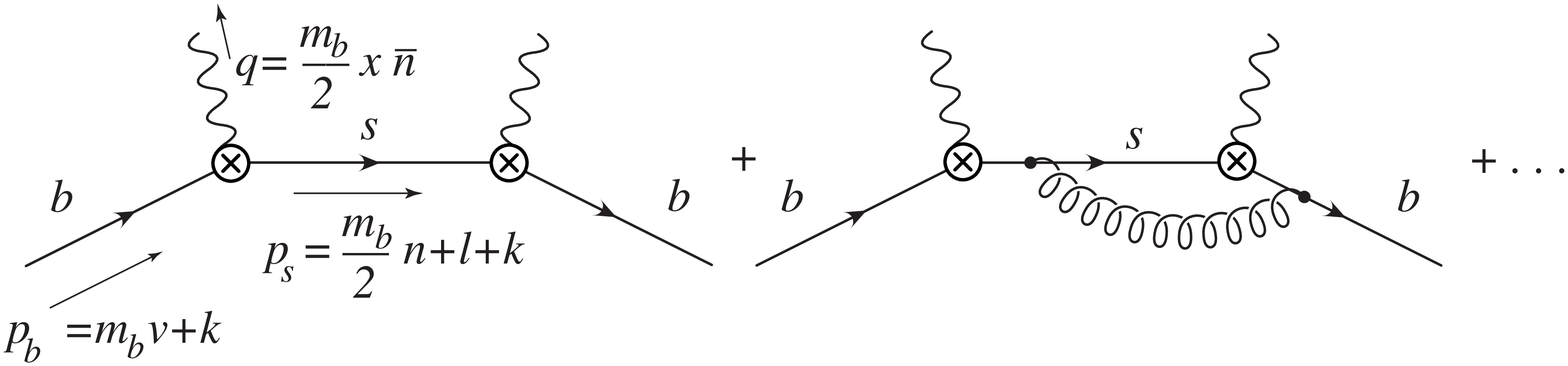}\hfill
     \caption{The OPE for $B\rightarrow X_s\gamma$.}
     \label{opegraphs}
\end{figure}

Consider the dominant contribution to the decay $B\rightarrow X_s
\gamma$, which arises from the magnetic penguin operator~\cite{gsw}
\begin{equation}
\hat{O}_7 = {e \over 16 \pi^2} m_b\; \bar{s} \, \sigma^{\mu \nu}
{1 \over 2} (1+ \gamma_5) b \; F_{\mu \nu} \,,
\label{o7}
\end{equation}
where the strange quark mass has been set to zero.\footnote{Throughout
  this work we will ignore the contribution of operators other than
  $\hat O_7$ to the decay.}  The OPE for this decay is illustrated in
Fig.\ \ref{opegraphs}.  We write the momenta of the $b$ quark, photon,
and light $s$ quark jet as
\begin{equation}
\label{kinematics}
p_b^\mu=m_b v^\mu+k^\mu,\ \ q^\mu={m_b\over 2} x \bar n^\mu,\ \ 
p_s^\mu={m_b\over 2} n^\mu + l^\mu + k^\mu
\end{equation}
where, in the rest frame of the $B$ meson,
\begin{equation}
v^\mu=(1,\vec 0),\ \ n^\mu=(1,0,0,-1),\ \ \bar
n^\mu=(1,0,0,1). 
\end{equation}
Here $k^\mu$ is a residual momentum of order $\Lambda_{\rm QCD}$, and
$l^\mu={m_b\over 2}(1-x) \bar n^\mu$, where $x = 2 E_\gamma/m_b$.  The
invariant mass of the light $s$-quark jet
\begin{equation}
p^2_s \approx m_b \, n \cdot (l+k) = m^2_b (1-x+\hat k_+)\,,
\end{equation}
(where $\hat k_+ = k_+/m_b$) is $O(m^2_b)$ except near the endpoint of
the photon energy spectrum where $x \to 1$. Inclusive quantities are
calculated via the OPE by taking the imaginary part of the graphs in
Fig.\ \ref{opegraphs} and expanding in powers of $k^\mu/\sqrt{p^2_s}$.
As long as $x$ is not too close to the endpoint, this is an expansion
in powers in $k^\mu/m_b$, which matches onto local operators.
This leads to an expansion for the photon energy spectrum as a
function of $x$ in powers of $\alpha_s$ and $1/m_b$~\cite{xxx}:
\begin{eqnarray}\label{qcdinclusive}
\frac{d\Gamma}{dx} &=& \Gamma_0 \left\{\left[1-\frac{\alpha_s C_F}{4
\pi} \left( 2 \log \frac{\mu^2}{m_b^2} +5 + \frac{4}{3} \pi^2 \right)
\right] \delta(1-x) \right.  \nonumber\\ && \;\;\;\;\;\;\;\;\;\left. +
\frac{\alpha_s C_F}{4\pi} \left[ 7 + x -2 x^2 - 2(1+x)\log(1-x) -
\left( 4 \frac{\log(1-x)}{1-x} + \frac{7}{1-x} \right)_+ \right]
\right. \nonumber \\
&& \;\;\;\;\;\;\;\;\ \left.+{1\over 
2m_b^2}\left[(\lambda_1-9\lambda_2)\delta(1-x)
-(\lambda_1+3\lambda_2)\delta^\prime(1-x)-{\lambda_1\over 
3}\delta^{''}(1-x)\right]\right\}
\nonumber \\
&& \;\;\;\;\;\;\;\;\;\;+O(\alpha_s^2, 1/m_b^3),
\label{fullrate}
\end{eqnarray}
where
\begin{equation}
\Gamma_0 = {G^2_F |V_{tb} V^*_{ts}|^2 \alpha |C_7(\mu)|^2 \over 32
\pi^4} m^5_b  \left[ \frac{m_b(\mu)}{m_b} \right]^2 \,,
\label{gamma0}
\end{equation}
and the subscript ``+" denotes the usual plus distribution,
\begin{eqnarray}
{1\over (1-x)_+}&\equiv& \lim_{\beta\rightarrow 0}\left\{{1\over 1-x} 
\theta(1-x-\beta)+
\log(\beta)\delta(1-x-\beta)\right\}\nonumber\\
\left({\log(1-x)\over (1-x)}\right)_+&\equiv& \lim_{\beta\rightarrow 
0}\left\{{\log(1-x)\over
1-x}
\theta(1-x-\beta)+
{1\over 2}\log^2(\beta)\delta(1-x-\beta)\right\} \,.
\end{eqnarray}
The parameters $\lambda_1$ and $\lambda_2$ are matrix elements of
local dimension five operators.

Near the endpoint of the photon spectrum, $x\rightarrow 1$, both the
perturbative and nonperturbative corrections are singular and the OPE
breaks down.  The severity of the breakdown is most easily seen by
integrating the spectrum over a region $1-\Delta < x < 1$.  When
$\Delta\leq \lqcd/m_b$ the most singular terms in the $1/m_b$
expansion sum up into a nonperturbative shape function of
characteristic width $\lqcd/m_b$\cite{shape}.  The perturbative series
is of the form
\begin{equation}
{1\over\Gamma_0}\int_{1-\Delta}^1 {d\Gamma\over dx}=1+{\alpha_s C_F\over 4
\pi}\left(-2
\log^2\Delta-7\log\Delta+\dots\right) + O \left( \alpha_s^2 \right),
\end{equation}
where the ellipses denote terms that are finite as $\Delta\rightarrow
0$.  These Sudakov logarithms are large for $\Delta\ll 1$, and can
spoil the convergence of perturbation theory. The full series has been
shown to exponentiate \cite{ks,ar} and the leading and next-to-leading
logarithms must be resummed for $\Delta\leq
\exp\left(-\sqrt{\pi/\alpha_s(m_b)}\right)$, which is parametrically
larger than $\lqcd/m_b$ in the $m_b\rightarrow \infty$ limit\cite{fjmw}.

{}In general, ``phase space'' logarithms are to be expected whenever a
decay depends on several distinct scales.  For example,
in $b\rightarrow X_c e\bar\nu_e$
decay the rate calculated with the OPE performed at $\mu=m_b$ contains large logarithms
of $m_c/m_b$.  In \cite{sumlogs} an EFT was used to 
run from $m_b$ to $m_c$, summing phase space logarithms of the ratio $m_c/m_b$.    Similarly,
in $b\rightarrow X_s\gamma$
near the endpoint of the photon energy spectrum the invariant mass of
the light quark jet scales as $m_b\sqrt{1-x}$, and is widely separated
from the scale $\mu=m_b$ where the OPE is performed.  In order to sum
logarithms of $\Delta$ (or the more complicated plus distributions in
the differential spectrum, Eq.\ (\ref{qcdinclusive})) we would expect
to have to switch to a new effective theory at $\mu=m_b$, use the
renormalization group to run down to a scale of order $m_b\sqrt{1-x}$,
at which point the OPE is performed.  (In fact, we will see that
the situation is slightly more complicated than this).

We are then left with the question of the appropriate theory below the
scale $m_b$.  The simplest possibility is to expand the theory in
powers of $k^\mu/m_b$ and $l^\mu/m_b$.  The heavy quark is then
treated in the heavy quark effective theory (HQET) \cite{hqetrefs},
while the light quark propagator is treated in the large energy
effective theory (LEET) proposed many years ago by Dugan and
Grinstein\cite{dg}.  Expanding the $s$ quark propagator in powers of
$1/m_b$, we find the LEET propagator
\begin{equation}
{i\xslash{p}_s\over p_s^2}={\xslash{n} \over 2}
{i\over n \cdot(l+k)}+O\left({l^\mu+k^\mu\over m_b}\right).
\end{equation}
LEET is an effective theory of lightlike Wilson lines, much as HQET is
an effective theory of timelike Wilson lines\cite{gk}.  The hope would
then be to match QCD onto LEET and then use the renormalization group
to sum the Sudakov logarithms.  This is the approach taken in \cite{bmk}.
However, a simple attempt at matching shows that this does not sum the
appropriate logarithms.

\begin{figure}[htbp]
  \epsfxsize=14cm \hfil\epsfbox{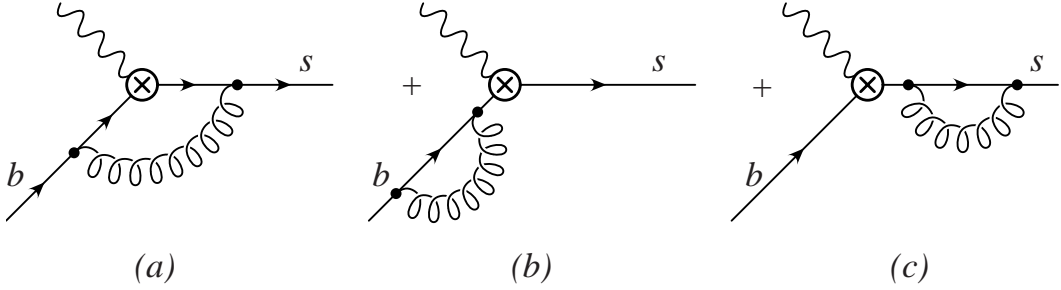}\hfill
     \caption{One-loop corrections to the matrix element of $\hat O_7$ 
     in QCD.}
     \label{v3}
\end{figure}

Consider the one-loop matching of the operator $\hat O_7$ from QCD to
LEET. We regulate ultraviolet (UV) divergences with dimensional
regularization ($d=4-2\epsilon$). We introduce a small invariant mass
$p^2_s$ for the $s$ quark which regulates all infrared (IR)
divergences except that in the heavy-quark wave function diagram,
Fig.\ \ref{v3}(b). This IR divergence is regulated using dimensional
regularization. The vertex diagram, Fig.\ \ref{v3}(a) yields
\begin{equation}
A^{(a)}_{QCD}= -C_7(\mu) \bar{s} \Gamma^\mu b \frac{\alpha_s C_F}{4 \pi}
   \left[ \log^2 {p_s^2 \over m_b^2} + 2 \log {p_s^2 \over m_b^2}  
\finite{+\pi^2}\right]\,,
\label{Aqcd2}
\end{equation}
where
\begin{equation}
\Gamma^\mu = 
{e \over 8 \pi^2} m_b \, \sigma^{\mu \nu} {(1+\gamma_5) \over 2} 
q_\nu \,.
\label{gammamu}
\end{equation}
$C_7(\mu)$ is the Wilson coefficient of $\hat O_7$ and the
dots denote (here and in the rest of the paper) finite terms which
are not logarithmically enhanced.
Including a factor of $\sqrt{Z}$ for each external field
\begin{eqnarray}
Z_b &=& 1-\frac{\alpha_s C_F}{4 \pi} \left[ \frac{3}{\epsilon} + 3
  \log \frac{\tilde \mu^2}{m_b^2}\finite{+4} \right] \label{wave_b} \\
Z_s &=& 1-\frac{\alpha_s C_F}{4 \pi} \left[ \frac{1}{\epsilon} - \log
  \frac{p_s^2}{m_b^2} + \log \frac{\tilde \mu^2}{m_b^2} \finite{- 1}
\right]\,,\label{wave_s} 
\end{eqnarray}
where
\begin{equation}
\tilde \mu^2 \equiv 4 \pi \mu^2 e^{-\gamma_E}
\end{equation}
and adding the counterterm required to subtract off the UV divergence
\begin{equation}
Z_7 = 1 + \frac{\alpha_s C_F}{4 \pi} \frac{1}{\epsilon}
\end{equation}
we find
\begin{equation}
A_{QCD}= C_7(\mu) \bar{s} \Gamma^\mu b \left[ 1-\frac{\alpha_s C_F}{4 \pi}
   \left( \log^2 {p_s^2 \over m_b^2} + \frac{3}{2} \log  {p_s^2 \over
       m_b^2} + \frac{1}{\epsilon} +
       2 \log {\tilde\mu^2 \over m_b^2} \finite{+ \pi^2 + \frac{3}{2}} 
       \right) \right]\,.
\label{Aqcd}
\end{equation}

\begin{figure}[htbp]
  \epsfxsize=4cm \hfil\epsfbox{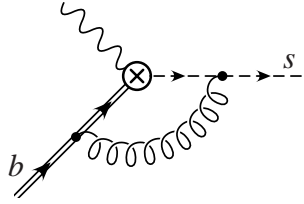}\hfill
     \caption{One-loop correction to the $bs\gamma$ vertex in LEET.}
     \label{leetfig}
\end{figure}

The corresponding LEET diagram is shown in Fig.\ \ref{leetfig}.
Neither of the wave function graphs gives a contribution, since the
light quark wave function in Feynman gauge\footnote{We will work in
Feynman gauge throughout this
paper.} is proportional to $n^2 = 0$, and
the heavy 
quark wave function vanishes in dimensional regularization. Thus the
only contribution is from the vertex graph.  Denoting the coefficient
of the corresponding operator in LEET as $C^{(0)}(1+(\alpha_s C_F/ 4\pi)
C^{(1)}+ \dots)$, we find
\begin{eqnarray}
A_{LEET}&=& C^{(0)}(\tilde\mu)\bar{\xi}_n \Gamma^\mu h \, \left[1-\frac{\alpha_s C_F}
{4 \pi} 
\left(\left( 4 \pi
    \, \frac{\mu^2 m_b^2}{p_s^4} \right) ^ {\!\epsilon} 
\frac{\Gamma(1+2\epsilon)
      \Gamma(1-2\epsilon) \Gamma(1+\epsilon)}{\epsilon^2}-C^{(1)}(\tilde\mu)
\right)\right] \nonumber \\ 
&=&C^{(0)}(\tilde\mu)\bar{\xi}_n \Gamma^\mu h \left[1-\frac{\alpha_s C_F}{4 \pi}
\left({1\over\epsilon^2}-{2\over\epsilon}\log{p_s^2\over m_b \tilde\mu} + 
2 \log^2 {p_s^2\over m_b \tilde\mu} 
\finite{+ \frac{3 \pi^2}{4}} - C^{(1)}(\tilde\mu)
\right)\right]\,,
\label{softterm}
\end{eqnarray}
where $p_s^2 /m_b$ is the soft scale. So
\begin{eqnarray}
C^{(0)}(\tilde\mu)&=&C_7(\tilde\mu)\nonumber\\
C^{(1)}(\tilde\mu)&=&{1\over \epsilon^2}-{1\over\epsilon}\left(2
\log{p_s^2\over m_b 
\tilde\mu}+1\right)+2\log^2{p_s^2\over m_b \tilde\mu}-\log^2{p_s^2\over m_b^2}
-{3\over2}\log{p_s^2\over m_b^2}\nonumber\\&&-2\log{\tilde\mu^2\over 
m_b^2}\finite{-
{\pi^2\over4}-{3\over 2}} .
\end{eqnarray}
We immediately notice two problems\footnote{Note that $C_7(\mu)$ includes 
a factor of $\alpha_s C_F/(4\pi) \log(m_W/\mu)$, which converts one of
the factors of $\log(\mu/m_b)$  in Eq.\ (\ref{Aqcd2}) to 
$\log(m_W/m_b)$.  This is not important for our argument.}:
\begin{enumerate}
\item There is no matching scale $\tilde\mu$ at which all the large single
  and double logarithms in $C^{(1)}$ vanish. Thus, there are logarithms in the
  rate which cannot be summed using the renormalization group in LEET.

\item $C^{(1)}$ contains a divergence proportional to
  $\frac{1}{\epsilon}\log p_s^2$.  Since $p_s^2$ is an infrared scale
  in the problem, it is not clear how to sensibly renormalize this
  term.   In Ref. \cite{bmk} this divergence was cancelled by 
  a nonlocal counterterm in the inclusive rate; however, this term
  indicates that LEET cannot be used for exclusive
  processes \cite{AC}.  Furthermore, the matching of the inclusive
  rate performed in \cite{bmk} still leaves large
  logarithms in the coefficient of the operator.
\end{enumerate}

The problem is that LEET only describes the coupling of light-like
particles to soft gluons, but does not describe the splitting of an
energetic particle into two almost collinear particles.  Thus, by
matching onto LEET one is integrating out the collinear modes which
also contribute to infrared physics. As we will show below, once
collinear degrees of freedom are included, both of the above problems
are resolved.

\section{The Collinear-Soft Theory}

\subsection{Collinear and soft modes}

It is convenient to work in light-cone coordinates $p^\mu =
(p^+,p^-,p^i_\perp)$, where $p^+ = n \cdot p$ and $p^- = \bar n \cdot
p$, and to define a power-counting parameter $\lambda = \sqrt{1-x}$
that becomes small in the limit $x \to 1$.  The momentum of the
light-quark jet then scales as
\begin{equation}
p_s^\mu\sim m_b(\lambda^2, 1, \lambda).
\end{equation}
This scaling is unchanged by emission of either soft or collinear
degrees of freedom, with momenta scaling as
\begin{equation}
p_{\rm soft}\sim m_b(\lambda^2, \lambda^2, \lambda^2),\ \ 
p_{\rm collinear}\sim m_b(\lambda^2, 1, \lambda),
\end{equation}
and so emission of both modes is kinematically allowed. It is the
presence of infrared sensitive graphs with collinear loop momentum
that makes this EFT more complicated than other, more familiar, EFT's,
where infrared sensitivity comes purely from soft modes.  This is
similar to the situation in non-relativistic QCD (NRQCD)\cite{nrqcd},
in which power counting is complicated by the fact that a given
amplitude receives contributions from loop momenta which are small
compared to the heavy quark mass, but which have parametrically
different dependence on the heavy quark velocity $v$.  In NRQCD, the relevant
scales are known as soft, ultrasoft and potential, and must be treated
separately in order to obtain consistent power
counting\cite{bs,rescale}.

We follow a similar approach here, and introduce separate fields for
both soft and collinear degrees of freedom\footnote{At two loops an
additional gluon field scaling as $(\lambda,\lambda,\lambda)$ might
have to be included~\cite{kor}.}.  Between the scales $m_b$ 
and $m_b\lambda$ the effective theory contains separate fields for
both collinear and soft modes, while at scales below $\sim
m_b\lambda$ (the exact scale depends on the operator under consideration,
as will be discussed in the next section), the collinear modes are 
integrated out of the theory and
it is matched onto LEET.  We will refer to this intermediate theory as
the collinear-soft theory, and resist the urge to create another
acronym. 

There is an important difference between the approach taken here and
the one taken in Refs.\cite{vnrqcd,ai} where logarithms of $v$ are
summed in NRQCD and NRQED. In the latter case no intermediate theory
is introduced; instead the running is performed in one step through
the velocity RGE. In NRQED these two approaches differ at subleading
order~\cite{ai,pc}, and it may be that such one-step running is needed
here at two loops. 

\begin{table}[tb]
\begin{tabular}{@{\hskip 1cm}c|c@{\hskip 4cm}}
Factor  &  Scaling    \\
\hline
soft gluon $A_\mu^s$ &  $\lambda^2$  \\
\hskip 2 cm  collinear gluon $A_\mu^c$\hskip 2 cm\  
&  $\lambda$    \\
heavy quark $h$ &  $\lambda^3$ \\
collinear quark $\xi$ & $\lambda$ \\
collinear volume element $d^4 x_c$ & $\lambda^{-4}$ \\
soft volume element $d^4 x_s$ & $\lambda^{-8}$ 
\end{tabular} 
\vspace{4pt}
\caption[]{
Power counting rules for fields in the collinear-soft theory
in Feynman gauge,
where $\lambda=\sqrt{1-x}$.
\label{tableone}}
\end{table}

The power counting rules in the collinear-soft theory may be obtained
by a field rescaling, analogous to that performed in \cite{rescale}.
The scaling of the fields is chosen such that the propagators are all
$O(1)$, putting the $\lambda$ dependence into the interaction terms.
For example, in the kinetic term for a soft gluon,
\begin{equation}
  \sim\int d^4 x\,(\partial_\mu A_\nu^a-\partial_\nu A_\mu^a)^2
\end{equation}
the typical length scale associated with soft excitations scales as
$\lambda^{-2}\sim p_{\rm soft}^{-1}$, so the factor of
$d^4x$ scales as $\lambda^{-8}$. Each derivative scales as
$p_{\rm soft}\sim \lambda^2$, so the soft gluon field must scale as
$\lambda^2$ for the kinetic term to be $O(1)$.  

Since the various collinear momentum components scale differently with 
$\lambda$, power counting for collinear gluons is gauge dependent (this is
easily seen from the propagator, since in a covariant gauge the
components of the $k^\mu k^\nu$ term scale differently). In this paper
we are working in Feynman gauge, in which case the different components of
collinear gluons have the same scaling.  Performing a similar analysis for the
other fields, we obtain the power-counting rules given in Table
\ref{tableone}.  

Rather than write down the effective Lagrangian for the various
fields, which is quite lengthy, we will instead just give the Feynman
rules, which are obtained by expanding the QCD amplitudes in powers of
$\lambda$.  The spinors in the collinear-soft theory are related, 
at leading order in $\lambda$, to the QCD spinors via
\begin{equation}
h_v = P_+ u \, , \,
\xi_n = P_n u \, , \,
\xi_{\bar{n}} = P_{\bar{n}} u \, , 
\label{eftspinors}
\end{equation}
where we have defined the projection operators
\begin{equation}
P_+ = \frac{\xslash{v}+1}{2} \, , \,
P_n = \frac{\xslash{n} {\xslash{\bar n}}}{4} \, , \,
P_{\bar{n}} =\frac{{\xslash{\bar n}} \xslash{n}}{4} \, ,
\label{hqetprojector}
\end{equation}
which project out the heavy quark spinor, a massless spinor in the $n$
direction, and a massless spinor in the $\bar{n}$ direction
respectively.
\begin{figure}[tbp]
  \epsfxsize=14cm \hfil\epsfbox{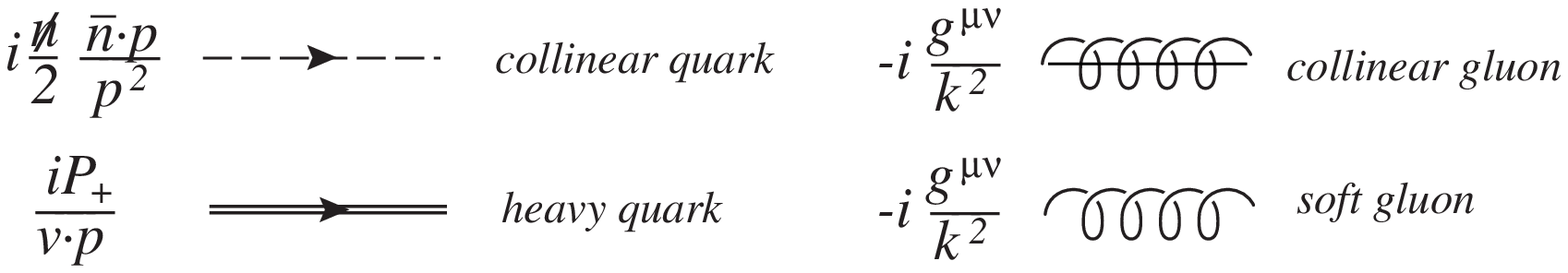}\hfill
     \caption{Propagators in the collinear-soft effective theory.}
     \label{v5}
\end{figure}
The propagators for the different fields are 
shown in Fig. \ref{v5}. 

\begin{figure}[hbt]
  \epsfxsize=14cm \hfil\epsfbox{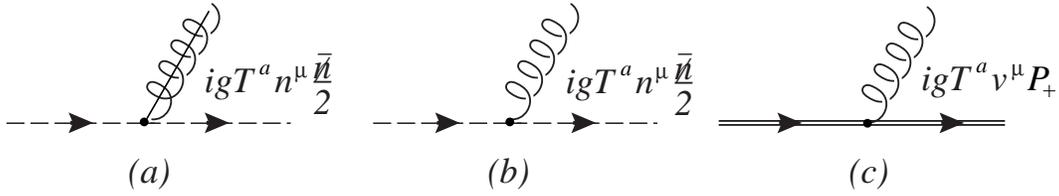}\hfill
     \caption{Leading order quark-gluon interactions in the collinear-soft 
       effective theory:  (a) collinear-collinear, (b) collinear-soft,
       and (c) heavy-soft.  Applying the rules from Table
       \ref{tableone}, the vertices scale as (a) $\lambda^{-1}$, (b)
       $\lambda^0$, and (c) $\lambda^0$.}
     \label{v2}
\end{figure}

The interactions leading in $\lambda$ which we will need in this
paper are shown along with
their Feynman rules and scaling in Fig.\ \ref{v2}.
Note that the interaction of a soft particle with a collinear particle
leaves the minus and perpendicular momenta of the collinear particle
unchanged, since they are parametrically larger for the collinear
particle.  This is analogous to the multipole expansion which is
performed in NRQCD\cite{nrqcd2}.  As a result at one loop,
soft-collinear interactions in this theory are equivalent to LEET,
since collinear propagators in soft loops reduce to LEET propagators:
\begin{equation}
\frac{\xslash{n}}{2} \; \frac{\bar n \cdot (p-k)}{(p-k)^2} \sim
\frac{\xslash{n}}{2} \; \frac{p^-}{(p-k)^+p^- -
    (p^\perp)^2} = -\frac{\xslash{n}}{2} \;
\frac{1}{n \cdot k}  \,
\label{eikonalprop}
\end{equation}
where $p$ is a collinear momentum, $k$ is a soft momentum, and $p^2=0$
from the equations of motion.  Once again, this is analogous to NRQCD,
where in ultrasoft loops the Feynman rules reduce to those for HQET.
By the same token, in soft-collinear interactions, the appropriate volume
element is the collinear volume element, scaling as $\lambda^{-4}$.

\begin{figure}[hbt]
  \epsfxsize=3.5 cm \hfil\epsfbox{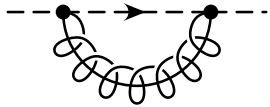}\hfill
\caption{Pure collinear wavefunction graph.  The $O(\lambda^{-2})$
contribution vanishes.}
\label{wfig}
\end{figure}

Because the leading purely collinear interaction, Fig. \ref{v2}(a),
scales as $\lambda^{-1}$,
power counting for collinear loops is less simple than for soft loops.
Terms which would scale as $\lambda^{-2}$, such as the purely collinear
wave-function graph in Fig.\ \ref{wfig}, are proportional to 
$n^2=0$ and so vanish in the effective 
theory.  
However, the $1/\lambda$ coupling enhances terms which would
na\"\i vely be suppressed.  
In fact, although the $\lambda$-counting looks complicated, graphs with
{\it only} collinear lines are identical to the corresponding graphs in
QCD.  This is because in any graph in which all the lines have the same
scaling (and there are no purely soft graphs, so this only refers to 
purely collinear graphs), expanding in powers of $\lambda$ does not change
the propagators.  Since the locations of 
poles in the propagators are unaffected, it is irrelevant whether one
calculates the full graph in QCD and then expands in powers of $\lambda$,
or calculates each order in $\lambda$ in the collinear-soft theory.
Thus, for 
purely collinear graphs, such as the wave function graph in
Fig.~\ref{wfig}, we will not bother to write down the complete
set of operators, but simply calculate the graph in QCD and expand.

\begin{figure}[bhtp]  
  \epsfxsize=10cm \hfil\epsfbox{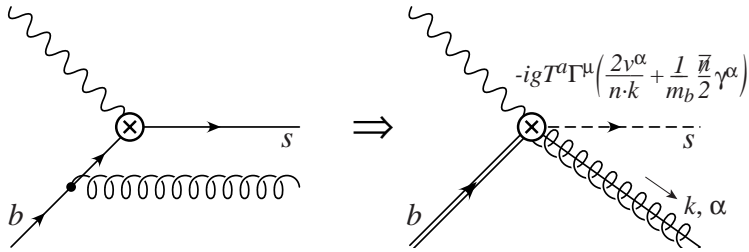}\hfill
\caption{Non-local vertex at $O(\lambda)$.  
\label{v1}}    
\end{figure}

There is one important subleading operator, shown
in Fig.\ \ref{v1}, which can be enhanced by the $1/\lambda$ piece of
the purely collinear coupling. By momentum conservation, there is no
vertex coupling two heavy quarks and a collinear quark, since a heavy
quark cannot emit a collinear gluon and stay on its mass shell. 
However, expanding the diagram in
Fig.\ \ref{v1} in powers of $\lambda$ gives the nonlocal $O(\lambda)$
interaction shown in the figure.  (This is similar to the nonlocal
operators found in \cite{benstuff}).
Though it is formally subleading, in graphs such as Fig. \ref{v4b}(a)
it gives an $O(1)$ effect. 

\subsection{Matching onto the Collinear-Soft Theory} 

\begin{figure}[bhtp]
  \epsfxsize=9cm \hfil\epsfbox{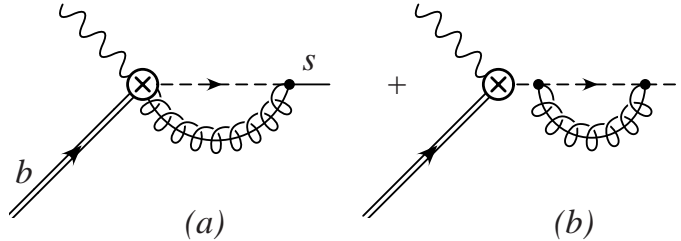}\hfill
\caption{The one-loop collinear gluon corrections to the vertex $V^\mu$.
\label{v4b}}
\end{figure}

\OMIT{The interactions given in the previous section suffice to
  compute the decay $b\rightarrow s\gamma$ at leading order in
  $\lambda$ in the collinear-soft theory, so we may now compute the
  matching conditions for the operator $\hat O_7$.}  We now proceed to
compute the matching conditions for the operator $\hat O_7$, and
demonstrate that there are no large logarithms in the matching
coefficients.  At tree level, the matching is trivial.  Defining the
current in the effective theory by
\begin{equation} 
V^\mu = \bar\xi_n \Gamma^\mu h_v \,, 
\label{colcurrent} 
\end{equation} 
where $\Gamma^\mu$ is given in (\ref{gammamu}), the Wilson coefficient
$C_V$ at tree level is
\begin{equation}  
C_V = 1 + O(\alpha_s). 
\label{cv} 
\end{equation} 
To perform this matching at one-loop, we repeat the one-loop matching
calculation discussed in Section \ref{scales}, but now using the
collinear-soft theory
instead of LEET, hence including collinear modes. The calculation is
simplest if we set the invariant mass of the $s$ quark to zero; this
introduces additional infrared divergences to the calculation which
cancel in the matching conditions. The one loop matrix element of
$\hat O_7$ in full QCD can be calculated from the diagrams in Fig.\ 
\ref{v3}, and we find the amplitude
\begin{equation}
A_{\rm QCD}=\bar s \Gamma^\mu b \left[1-\frac{\alpha_s C_F}{4 \pi} 
\left(
    \frac{1}{\epsilon^2} + \frac{\log(\tilde\mu^2/m_b^2)}{\epsilon} +
    \frac{5}{2\epsilon} + \frac{1}{2} \log^2 \frac{\tilde
    \mu^2}{m_b^2} + \frac{7}{2} \log \frac{\tilde\mu^2}{m_b^2}\finite{ +
  \frac{\pi^2}{12} + 6} \right)\right] \,.
\label{qcddiv}
\end{equation}
where all the $1/\epsilon$ divergences are infrared in origin.
The one loop correction in the collinear-soft theory can be calculated
from the Feynman 
diagrams in Figs.\ \ref{leetfig} and \ref{v4b}. In pure dimensional
regularization all graphs are zero, as there is no scale present in
the loop integrals.  Thus, we find the matching condition
\begin{equation}\label{eff_counter}
C_V\,Z_V = \finiteomit{\left[1 -  \frac{\alpha_s C_F}{4 \pi}
  \left(\frac{\pi^2}{12} + 6 \right) \right] \left[}1 + \frac{\alpha_s
    C_F}{4 \pi} \left( 
    \frac{1}{\epsilon^2} + \frac{\log(\tilde\mu^2/m_b^2)}{\epsilon} +
    \frac{5}{2\epsilon} \finite{}\right)  \finiteomit{\right]}\,,
\label{counterterm}
\end{equation}
where $Z_V$ is the counterterm required to subtract the UV divergences
in the collinear-soft theory. 

This derivation of course assumes that the collinear-soft theory reproduces the
infrared behaviour of QCD.  We can check this by instead introducing a
small invariant mass for the $s$ quark, as in Section \ref{scales},
and explicitly verifying that the dependence on the invariant mass in
the collinear-soft theory is identical to that in full QCD given in Eq.~(\ref{Aqcd}).
The soft gluon contribution in the collinear-soft theory is identical to the LEET result,
given in (\ref{softterm})
\begin{eqnarray}
A_s &=& - C_V \bar{\xi}_n \Gamma^\mu h \frac{\alpha_s C_F}{4 \pi}
\left[{1\over\epsilon^2}-{2\over\epsilon}\log{p_s^2\over m_b \tilde\mu} + 
2 \log^2 {p_s^2\over m_b \tilde\mu} \finite{}
\right]\,.
\label{softterm2}
\end{eqnarray}
The collinear vertex diagram, Fig.\ \ref{v4b}(a), gives
\begin{eqnarray}
A^{(v)}_c&=& C_V \bar{\xi}_n \Gamma^\mu h \frac{\alpha_s C_F}{2 \pi} 
    \left( 4 \pi  \frac{\mu^2}{p_s^2} \right) ^ {\!\epsilon}  
    \frac{\Gamma(1+\epsilon)\Gamma(1-\epsilon)\Gamma(2-\epsilon)}
    {\Gamma(2-2 \epsilon)}
    {1 \over \epsilon^2} \nonumber\\
&=& - C_V \bar{\xi}_n \Gamma^\mu h \frac{\alpha_s C_F}{4 \pi} \left[
    -\frac{2}{\epsilon^2} - \frac{2}{\epsilon} + \frac{2}{\epsilon}
    \log\frac{p_s^2}{\tilde\mu^2} - \log^2{p_s^2\over \tilde\mu^2} 
    + 2 \log{p_s^2\over \tilde\mu^2} 
    \finite{+ \frac{\pi^2}{6} - 4}\right]\,.
\label{collinearvertex}
\end{eqnarray}
As previously discussed, the leading piece of the wave function graph
Fig.\ \ref{v4b}(b) is $O(1/\lambda^2)$, but fortunately vanishes.
Expanding to higher orders in $\lambda$, the graph gives the same
result as in full QCD, (\ref{wave_s}).  We therefore obtain for the
contribution of the collinear gluons
\begin{eqnarray}
A_c &=& - C_V \bar{\xi}_n \Gamma^\mu h \frac{\alpha_s C_F}{4 \pi} 
    \left[-{2 \over \epsilon^2} - {3 \over 2 \epsilon} + 
    {2 \over \epsilon} \log{p_s^2\over \tilde\mu^2} - 
    \log^2{p_s^2\over \tilde\mu^2} + {3 \over 2} \log {p_s^2\over \tilde\mu^2} 
    \finite{+ \frac{\pi^2}{6} - \frac{9}{2}}
\right]\,.
\label{collinearterm}
\end{eqnarray}

Adding the soft and collinear contributions, as well as the
counterterm given in (\ref{counterterm}), we obtain
\begin{equation}
A_{cs} = - C_V \bar{\xi}_n \Gamma^\mu h \frac{\alpha_s C_F}{4 \pi} \left[
   \log^2{p_s^2\over m_b^2} + \frac{3}{2} \log {p_s^2\over m_b^2} + {1 \over
   \epsilon} -\frac{1}{2} \log^2 {\tilde\mu^2\over m_b^2} - 
   {3 \over 2}\log {\tilde\mu^2\over m_b^2} 
   \finite{+ \frac{11 \pi^2}{12}  - \frac{9}{2}} \right] \,.
\end{equation}
Note that the troublesome divergence $\sim \frac{1}{\epsilon} \log
p_s^2$ cancels once the two contributions (\ref{softterm2}) and
({\ref{collinearterm}) are added. Thus, both collinear and soft modes
  are required for the theory to be renormalized sensibly.  Comparing
  to the full theory result (\ref{Aqcd}), we see that the
  collinear-soft theory reproduces 
  the IR physics of QCD, and that at the scale $\tilde\mu = m_b$ all
  nonanalytic terms vanish.  This determines the matching scale to be
  $m_b$,\finiteomit{and the matching coefficient at one loop is
\begin{equation}
C_V(m_b) = 1 - \frac{\alpha_s C_F}{4 \pi} \left[\frac{\pi^2}{12} + 6
\right]\,,
\end{equation}
}
confirming the result found by calculating in pure dimensional
regularization (\ref{eff_counter}).  \OMIT{
%
  {}It is instructive to calculate the differential decay rate
  $d\Gamma/dx$ in the collinear-soft theory. By the optical theorem the differential decay
  rate is given by the imaginary part of the forward scattering matrix
  element
\begin{equation}
\frac{d\Gamma}{dx} = {\rm Im}\, T(x)\,.
\end{equation}
The Feynman diagrams for the forward scattering matrix element are
shown in Fig.\ \ref{rescaledfs}.
\begin{figure}[htbp]
  \epsfxsize=14cm \hfil\epsfbox{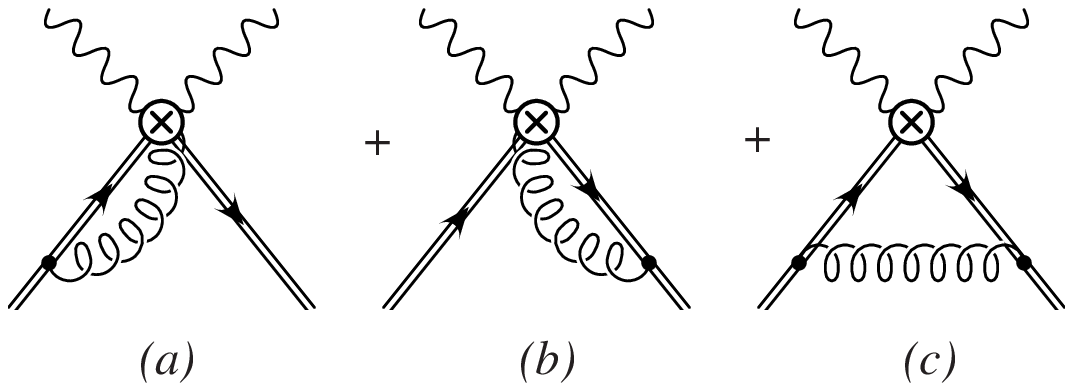}\hfill
     \caption{Collinear--soft theory Feynman diagrams which contribute
       to the forward scattering amplitude through $O(\alpha_s)$.}
     \label{rescaledfs}
\end{figure}
and we regulate all divergences in dimensional regularization. Since
we are considering an inclusive quantity, the result is free from
infrared divergences.  Thus all poles in $\epsilon$ are UV and are
canceled by the counterterm $Z_V$ for each $bs\gamma$ vertex.  Since
we are interested in the differential decay rate close to the endpoint
$x=1$, we expand the expression for the forward scattering amplitude
obtained from these graphs in powers of $(1-x)$ and find
\begin{eqnarray}
\label{forward}
T(x) &=& \lim_{\eta\to 0} \left(-\frac{1}{\pi}\right) \frac{C_V^2(\mu)
  \Gamma_0 }{1-x+i 
  \eta} \Bigg\{ 1 + 
  \frac{\alpha_s C_F}{4 \pi} \Bigg(\log^2 \frac{\mu^2}{m_b^2} + 5 \log
  \frac{\mu^2}{m_b^2} \nonumber\\
&&\;\;
- 2 \log^2 (1-x) + 4 i \pi \log(1-x) - 7
  \log(1-x) + 7 i \pi \Bigg)\Bigg\} + {\rm analytic}\,.
\end{eqnarray}
This leads to the differential decay rate
\begin{eqnarray}
\label{diff_LCET}
\frac{d\Gamma}{dx} &=& C_V^2(\mu) \Gamma_0 \Bigg\{ \left[ 1 +
  \frac{\alpha_s C_F}{4 \pi} \left(\log^2 \frac{\mu^2}{m_b^2} + 5 \log 
  \frac{\mu^2}{m_b^2} \right)\right] \delta(1-x) \nonumber\\
&&\qquad\qquad\qquad
-  \frac{\alpha_s C_F}{4 \pi} \left[ 4 \left(\frac{\log(1-x)}{1-x}\right)_+ + 7
  \left(\frac{1}{1-x}\right)_+ \right] \Bigg\}\,.
\end{eqnarray}
Comparing to (\ref{fullrate}), we can see that the effective theory
reproduces all the nonanalytic dependence on $1-x$ of full QCD. Thus,
at the scale $\mu = m_b$ the difference between the collinear-soft theory and full QCD
contains no nonanalytic terms, confirming our earlier determination of
the matching scale.
\OMIT{ Note that in order to reproduce the plus-distributions
  correctly, the terms proportional to $i \pi$ in Eq.~(\ref{forward})
  are essential. It is convenient to absorb these terms into the
  logarithms by writing
\begin{eqnarray}
T(x) &=& \lim_{\eta\to 0} \left(-\frac{1}{\pi}\right) \frac{C_V^2(\mu)
  \Gamma_0 }{1-x+i 
  \eta} \Bigg\{ 1 + 
  \frac{\alpha_s C_F}{4 \pi} \Bigg(\log^2 \frac{\mu^2}{m_b^2} + 5 \log
  \frac{\mu^2}{m_b^2} \nonumber\\
&&\quad
- 2 \log^2 (x-1-i\eta) - 7
  \log(x-1-i\eta) \Bigg)\Bigg\} + {\rm analytic}\,.
\end{eqnarray}
This allows us to write the differential decay rate as
\begin{eqnarray}\label{LCET_diff}
\frac{d\Gamma}{dx} &=& C_V^2(\mu) \Gamma_0 \Bigg\{ \left[ 1 +
  \frac{\alpha_s C_F}{4 \pi} \left(\log^2 \frac{\mu^2}{m_b^2} + 5 \log 
  \frac{\mu^2}{m_b^2} \right) \right] \delta(1-x) \nonumber\\
&&\quad
-  \frac{\alpha_s C_F}{4 \pi} \left[ \frac{2}{\pi} {\rm Im} \left\{ 
     \frac{\log^2(x-1-i\eta )}{x-1-i\eta}
   \right\} + \frac{7}{\pi} {\rm Im} \left\{
     \frac{\log(x-1-i\eta)}{x-1-i\eta} \right\} \right] \Bigg\}\,.
\end{eqnarray}
This agrees with the previous result (\ref{diff_LCET}), since the plus
functions can be written as
\begin{eqnarray}
\frac{1}{(1-x)_+} &=& \lim_{\eta \to 0} \frac{1}{\pi} {\rm Im} \left\{
     \frac{\log(x-1-i\eta)}{x-1-i\eta} \right\} \\
\left( \frac{\log(1-x)}{1-x} \right)_+ &=& \lim_{\eta \to 0} 
\frac{1}{2\pi} {\rm Im}\left\{
     \frac{\log^2(x-1-i\eta )}{x-1-i\eta}
   \right\} + {\pi^2 \over 2} \delta(1-x) \,.
\end{eqnarray}
We will see later that by absorbing the terms proportional to $i\pi$
into the logarithms will allow us to resum plus distributions using
the standard RGE.  } }

\subsection{Renormalization group equations}

{}From the counterterm given in (\ref{counterterm}) it is simple to
extract the anomalous dimension of the operator $V^\mu$ in the collinear-soft theory. From
the definition,
\begin{equation}
 \gamma_V = Z^{-1}_V \left( \tilde\mu {\partial \over \partial \tilde\mu} +
\beta {\partial \over \partial g} \right) Z_V
\end{equation}
we have
\begin{eqnarray}
\tilde\mu{\partial\over\partial \tilde\mu}Z_V&=&{\alpha_s(\tilde\mu)
  C_F\over2\pi\epsilon}\nonumber\\ 
\beta{\partial\over\partial g}Z_V&=&-{\alpha_s(\tilde\mu) C_F\over
  2\pi}\left({1\over\epsilon}+\log{{\tilde\mu^2 \over 
m^2_b}} + {5 \over 2} \right)\,,
\end{eqnarray}
where we have used $\beta = - g \epsilon + O(g^3)$. This give the
anomalous dimension 
\begin{equation}\label{gamma_V}
\gamma_V=-{\alpha_s(\tilde\mu)C_F\over 2\pi}\left(\log{{\tilde\mu^2 \over
m^2_b}} + {5 \over 2} \right).
\end{equation}
Note that the divergent piece of the anomalous dimension cancels
between the two terms \cite{bmk}.  The RGE for the coefficient of the
operator $V^\mu$ is therefore
\begin{equation}
\tilde\mu {d \over d \tilde\mu} C_V (\tilde\mu) = \gamma_V(\tilde\mu) C_V (\tilde\mu) \,.
\label{collrge}
\end{equation}
Solving this RGE we obtain
\begin{equation}
C_V( \tilde\mu ) = \left( {\alpha_s( \tilde\mu ) \over \alpha_s}
\right)^{{C_F \over 2 \beta_0}\left(5 - {8 \pi \over \beta_0 
\alpha_s}\right)}
\left({\tilde\mu^2 \over m^2_b}\right)^{-{C_F \over \beta_0}} C_V(m_b) \,,
\label{cvmu}
\end{equation}
where $\alpha_s \equiv \alpha_s(m_b)$, $\beta_0 = 11-2/3 n_f$, and
$C_V (m_b) = 1+O(\alpha_s(m_b))$. Note that in deriving the anomalous
dimension (\ref{gamma_V}) we have assumed that the nonlocal vertex
given in Fig.~\ref{v1} has the same running as the QCD coupling.  
This assumption needs to be checked in subsequent work.
\OMIT{ It is instructive to expand out the solution to the RGE:
\begin{equation}
C_V(\mu)=1-{\alpha_s C_F \over 8\pi}\left(\log^2
\left({m_b^2\over \mu^2}\right)-
5\log\left({m_b^2\over\mu^2}\right)\right)+O(\alpha_s^2).
\end{equation}
As expected, at the one-loop level the $\mu$ dependence of the
coefficient $C_V$ cancels the $\mu$ dependence of the differential
decay rate in the collinear-soft theory, given in Eq.~(\ref{diff_LCET}). Note, however,
that there are still unsummed logarithms in the plus-distributions in
(\ref{diff_LCET}). These logarithms can be summed by matching onto a
different effective theory and evolving this new theory to a lower
scale. } }

\section{The soft theory}

\subsection{Matching}

\begin{figure}[bt]
  \epsfxsize=14cm \hfil\epsfbox{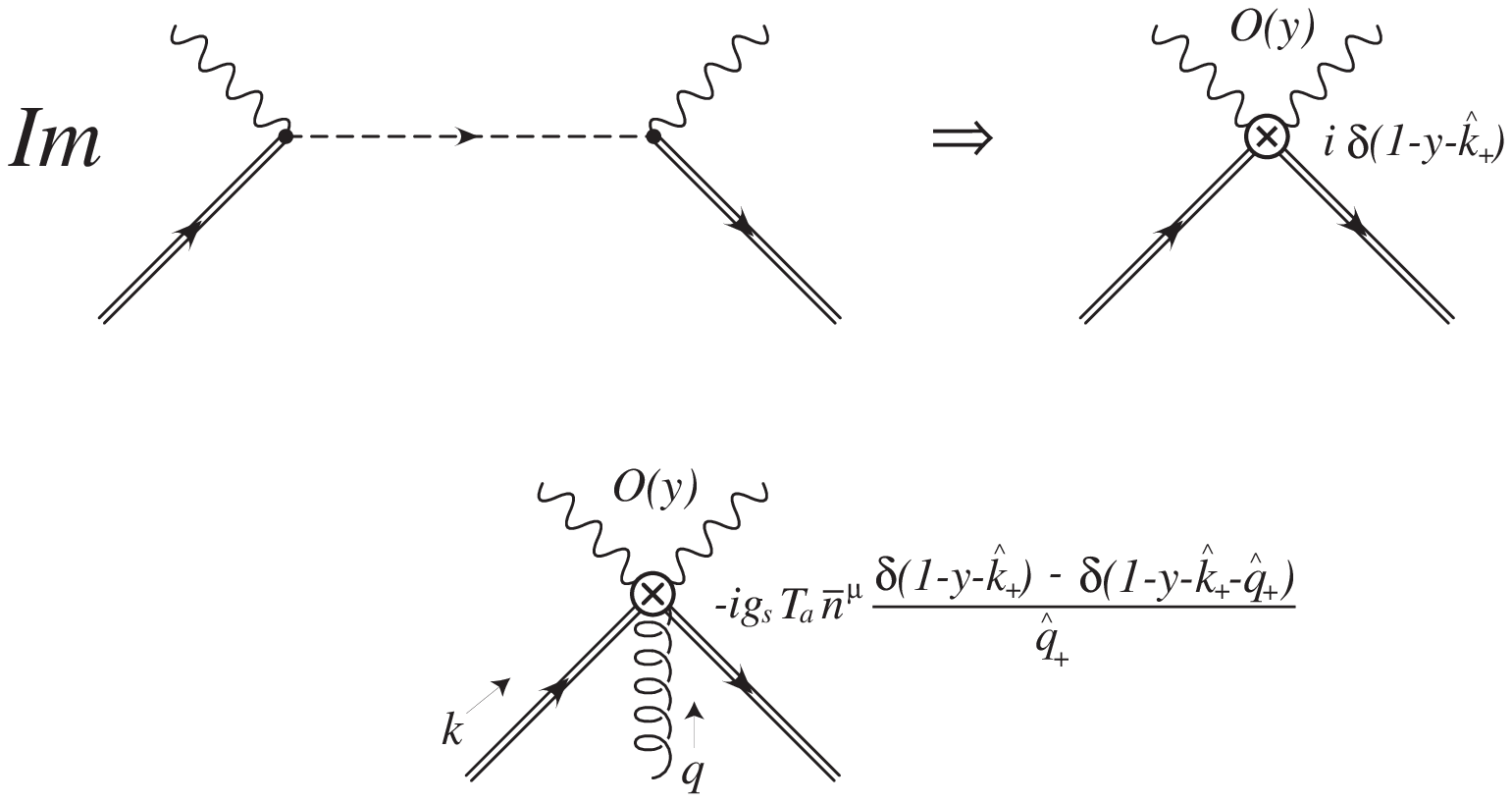}\hfill
     \caption{Diagrammatic representation of the OPE, as well as the zero and
     one gluon Feynman rules for the resulting nonlocal operator $O(y)$.}
     \label{leetopfeynrules}
\end{figure}

{}The collinear-soft effective theory is valid down to $\tilde\mu \approx
m_b \sqrt{1-x}$, the typical invariant mass of the light $s$-quark
jet.  At this scale we integrate out the collinear modes, and perform 
an OPE to calculate the inclusive $b$ decay rate.  Diagrammatically, this
is illustrated in Fig.\ (\ref{leetopfeynrules}).  This results in a nonlocal OPE in which the two currents are
separated along a light-like direction.  
As in Eq.\ (\ref{kinematics}), we write the momentum of the eikonal
line as 
\begin{equation}
p_s^\mu=\frac{m_b}{2} n^\mu+k^\mu+\frac{m_b}{2}(1-y)\bar n^\mu
\end{equation}
where $k^\mu$ is the residual momentum of the heavy quark (note that we 
distinguish $y$ from $x$, the rescaled photon momentum, since beyond tree 
level they will differ).  The imaginary piece of the first graph is then 
proportional to $\delta(1-y+\hat k_+)$ (where, as usual, hatted variables
are divided by $m_b$), so the OPE is in terms of an 
infinite number of nonlocal operators, labelled by $y$:
\begin{equation}
O(y) = \bar h_v  \, \delta(1 - y + i \hat{D}_+) \, h_v \,.
\label{LEETop}
\end{equation}
Feynman rules for nonlocal operators of this type were obtained in \cite{gn},
by writing them as the Fourier transform of operators in position space,
and expanding out the path-ordered exponential in powers of the gauge field.
Equivalently, the Feynman rules may be obtained by taking the imaginary piece
of the time-ordered product in LEET with additional gluons;  the
single gluon
Feynman rule is given in Fig.~\ref{leetopfeynrules}. 

The matrix element of $O(y)$ between heavy quark states with residual
momentum $k$ is
\begin{equation}
\langle b(k)|\bar h_v \delta(1-y+i\hat{D}_+)
    h_v | b(k) \rangle=\delta(1-y+\hat k_+)+O(\alpha_s)\,.
\end{equation}
while its matrix element between hadrons is the well known 
structure function \cite{shape}
\begin{equation}
f(y) = 
\frac{\langle B|\bar h_v \delta(1-y+i\hat{D}_+)
    h_v | B \rangle}{\langle B|\bar h_v h_v | B \rangle}\,.
\end{equation}
Thus, LEET consists of a continuous set of operators labeled by $y$.
Each operator has a coefficient that depends on the kinematic variable
$x$, and the differential rate for $B \to X_s \gamma$ is given by the
integral
\begin{equation}
\frac{d\Gamma}{dx} = \Gamma_0 \int \!dy \, C(y,x;\mu) f(y;\mu)\,,
\end{equation}
where the $C(y,x;\mu)$'s are the coefficients of the OPE. 

\begin{figure}[bh]
  \epsfxsize=9cm \hfil\epsfbox{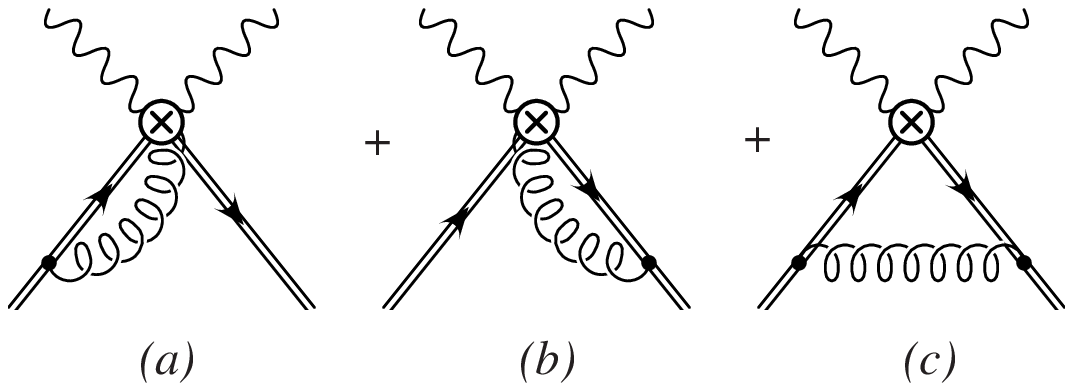}\hfill
     \caption{Feynman diagrams contributing to the one-loop matrix
       element of $O(y)$.}
     \label{leetgraphs}

\end{figure}

To match onto LEET at one loop we compare the differential decay rate
in the parton model, $b \to X_s \gamma$, which in LEET is
\begin{equation}
\label{diff_convolution}
\frac{d\Gamma}{dx}\Bigg|_{k_+} = \Gamma_0 \int \!dy \, C(y,x;\mu) \langle
b(k)| O(y;\mu) | b(k) \rangle\,.
\end{equation}
We therefore need the one-loop matrix element of $O(y)$ between quark
states. This may be calculated from the diagrams shown in
Fig.~\ref{leetgraphs}.

Again all divergences are regulated in dimensional regularization.  As
an example, Fig.\ \ref{leetgraphs}(a) gives
\begin{equation}
\langle b(k)|O^{(a)}(y)|b(k)\rangle= i C_F g^2 \left({\mu\over
m_b}\right)^{4-d}\int{d^{d-2}
\hat q_\perp\over(2\pi)^{d-2}} {d\hat q_-\over 2\pi}{d \hat q_+\over
2\pi} {\delta(\hat 
k_++1-y)-\delta(\hat k_+ + \hat q_++1-y)\over (\hat q_+
\hat q_--\hat q_\perp^2+i\epsilon)(\hat q_++\hat q_-+i\epsilon) \hat q_+}.
\end{equation}
The first term is proportional to
\begin{eqnarray}
&&\int {d^d \hat q \over (2\pi)^d} {1\over (\hat q^2+i\epsilon)(\hat q\cdot
  v+i\epsilon)(\hat q\cdot n)}
=
8\int {d^d \hat q\over(2\pi)^d}\int_0^1 dx\int_0^\infty d\lambda\,{\lambda\over
\left(\hat q^2+2\lambda \, \hat q\cdot(v(1-x)+x \, n)\right)^3}\nonumber \\
&&\qquad =-{4 i\over(4\pi)^{d/2}}\Gamma(3-d/2)\int_0^1 dx\int_0^\infty
d\lambda\, \lambda^{d-5} 
\left((1-x)^2+2(1-x)\right)^{d/2-3}.
\end{eqnarray}
The $\lambda$ integral vanishes in dimensional regularization, so
this term vanishes.  After performing the trivial $\hat q_+$ integral
in the second term, we are left with
\begin{eqnarray}
\langle b(k)|O^{(a)}(y)|b(k)\rangle&=&i\,{C_Fg^2\over 2\pi}\left({\mu\over
m_b}\right)^{4-d}{1\over
\hat k_++1-y} \int {d^{d-2}
\hat q_\perp\over (2\pi)^{d-2}} {d \hat q_-\over 2\pi} {1\over
\hat q_-(\hat k_++1-y)+\hat q_\perp^2-i\epsilon}\nonumber\\
&&\hskip 2.3in \times {1\over \hat q_--(\hat k_++1-y)+i\epsilon}
\nonumber \\
&=&{C_F g^2 \over 2\pi}\left({\mu\over m_b}\right)^{4-d}{\theta(\hat
  k_++1-y)\over \hat  
k_++1-y}\int {d^{d-2}\hat q_\perp\over (2\pi)^{d-2}} {1\over \hat q_\perp^2+
(\hat 
k_++1-y)^2}\nonumber\\ 
&=&{C_F g^2\over 8 \pi^2}\left(4\pi \,{\mu^2\over m_b^2}\right)^{\epsilon}
\Gamma(\epsilon){\theta(\hat k_++1-y)\over (\hat k_++1-y)^ {1+2\epsilon}}\,.
\end{eqnarray}
Using the identity
\begin{equation}
{\theta(y-x)\over(y-x)^{1+2\epsilon}}=-{1\over 2\epsilon}\delta(y-x)+
\theta(y-x)\left[{1\over
(y-x)_+}- 2\epsilon\left(\log(y-x)\over (y-x)\right)_++O(\epsilon^2)\right]
\end{equation}
we find
\begin{eqnarray}
\langle b(k)|O^{(a)}(y)|b(k)\rangle&=&-\,{\alpha_s C_F\over
4\pi}\left\{\left({1\over\epsilon^2}+{1\over\epsilon}\log{\tilde\mu^2
\over m_b^2}+{1\over 2}\log^2{\tilde\mu^2\over m_b^2} \finite{+{\pi^2\over
12}}\right)\delta(1-y+\hat k_+)
\right.\nonumber \\
&&-\theta(1-y+\hat k_+ ) \left[\left.\left({2\over\epsilon}+2\log{\tilde\mu^2\over m_b^2}\right)
{1\over (1-y+\hat k_+)_+}\right.\right.\nonumber\\
&&\hskip 1.5in\left.\left.-
4\left(\log(1-y+\hat k_+)\over 1-y+\hat k_+\right)_+\right]\right\}\,.
\end{eqnarray}

The diagram in Fig.\ \ref{leetgraphs}(b) gives the same result as (a),
while the diagram in Fig.\ \ref{leetgraphs}(c) gives
\begin{equation}
\langle b(k)\vert O^{(\rm c)}(y) |b(k)\rangle =  -\,\frac{\alpha_s
      C_F}{4 \pi}
\left[ \left( -{2 \over \epsilon} -2 \log {\tilde\mu^2 \over m_b^2} \right)
  \delta(1-y+\hat k_+)  + 4 \,\frac{\theta(1-y+\hat k_+)}
{(1-y+\hat k_+)_+} \right].
\end{equation}
In dimensional regularization the wavefunction diagrams vanish.  Since
the decay rate is infrared finite, including the wavefunction graphs
simply converts an infrared $1/\epsilon$ divergence to an ultraviolet
divergence.  Therefore, we may neglect the wavefunction counterterm,
and combining all graphs we find the bare matrix element
\begin{eqnarray}
\langle b(k)\vert O^{\rm{bare}}(y) \vert b(k)\rangle&=&\left[1-{\alpha_s C_F
\over 4\pi}\left({2\over\epsilon^2}-{2\over\epsilon}+{2\over\epsilon}
\log{\tilde\mu^2
\over m_b^2}-2\log{\tilde\mu^2\over m_b^2} \right.\right. \nonumber\\
&&\hskip 1.4in \left.\left.+\log^2{\tilde\mu^2\over
m_b^2} \finite{+ \frac{\pi^2}{12}} \right) \right] \delta(1-y+\hat k_+)
\nonumber\\ 
&&
+{\alpha_s C_F
\over 4\pi}\,\theta(1-y+\hat k_+)
\left[\left(-{4\over\epsilon}-4\log{\tilde\mu^2\over 
    m_b^2}+4\right) 
\frac{1}{(1-y+\hat k_+)_+}\right.\nonumber\\
&&\hskip 1.4in+\left.8\left({\log(1-y+\hat k_+)\over 1-y+\hat
k_+}\right)_+\right]\,,
\end{eqnarray}
where all divergences are ultraviolet.  The renormalized operator
$O(y;\mu)$ is related to the bare operator by
\begin{equation}
O^{\rm bare}(y) = \int \! dy' \,Z(y',y;\tilde\mu) O(y';\tilde\mu) \,.
\label{Orenormalization}
\end{equation}
Renormalizing in MS (generalized in the obvious way to cancel the
$1/\epsilon^2$ divergences), we find
\begin{eqnarray}
Z(y',y;\tilde\mu) &=& \left\{ \left[ 1 - {\alpha_s(\tilde\mu) C_F \over 2\pi}
        \left( {1\over  \epsilon^2} + {1 \over \epsilon} \log{\tilde\mu^2
        \over m_b^2} - {1 \over  \epsilon} \right) \right]\right.
        \delta(y'-y)
\nonumber \\
&& \qquad\qquad\qquad\qquad\left. 
+ {\alpha_s(\tilde\mu) C_F \over \pi} {1 \over
        \epsilon} \frac{1}{(y'-y)_+} \theta(y'-y) \right\}\,.
\label{softct}
\end{eqnarray}
\OMIT{and
\begin{eqnarray}
\langle b(k) | O(y;\mu)|b(k) \rangle &=&  \int dy'\,
\left\{ \delta(y'-y) -\frac{\alpha_s
      C_F}{4 \pi}
\left[ \left( \log^2 {\mu^2 \over m_b^2} - 2 \log {\mu^2 \over m_b^2}
  \right) \delta(y'-y) \right.\right.\nonumber\\
&& \left.\left. + \left( 4 - 4 \log {\mu^2 \over m_b^2} \right)
    \frac{1}{(y'-y)_+}  + 8 
\left(\frac{\log(y'-y)}{y'-y} \right)_+ \right] \right\}+ {\rm analytic} \,.
\label{leetresult}
\end{eqnarray}
}%
Note that the counterterm consists of a diagonal piece which is
proportional to $\delta(y'-y)$, and an off-diagonal piece proportional
to $\theta(y'-y)$. This latter terms mixes the operator $O(y)$ with
all operators $O(y')$ with $y' > y$.

Inserting the one-loop matrix element of the renormalized operator
into (\ref{diff_convolution}) we find the the differential decay rate
in the parton model $b \to X_s \gamma$
\begin{eqnarray}\label{leetrate}
\frac{d\Gamma}{dx} \Bigg|_{k_+}  &=& \Gamma_0 \int \!dy \, C(y,x;\tilde\mu) \langle O(y;\tilde\mu)
\rangle
\label{ope1}\nonumber\\
&=&\Gamma_0 \int \!dy \, C(y,x;\tilde\mu) \left\{\left[1-{\alpha_s C_F\over 4
\pi}\left(
\log^2{\tilde\mu^2\over m_b^2}-2\log{\tilde\mu^2\over m_b^2} 
\finite{+ \frac{\pi^2}{12}} \right) 
\right]\delta(1-y+k_+)
\right.\nonumber\\
&&\hskip .5in \left.
-{\alpha_s C_F\over 4\pi}\,\theta(1-y+\hat k_+) \left[\left(4-4\log{\tilde\mu^2\over m_b^2}\right)
{1\over(1-y+\hat
k_+)_+}\right.\right.\nonumber\\
&&\hskip 2.6in \left.\left.+8\left({\log(1-y+\hat k_+)\over 1-y+\hat
        k_+}\right)_+\right] \right\} \,.
\end{eqnarray}
One might worry about the appearance in (\ref{leetrate}) of
logarithmic terms that depend on $m_b$, since this scale has been
integrated out and thus should not be present in the effective
theory. These terms are due to our choice of factoring the
heavy quark mass out of the soft scale $m_b (1-y+ \hat k_+)$ by
writing our expressions in terms of hatted quantities. The logarithms of
$m_b$ cancel in the matching coefficient.

The Wilson coefficients $C(y,x;\mu)$ are determined by matching
the collinear-soft theory onto LEET.  In the collinear-soft theory,
the Feynman diagrams for the forward scattering matrix element are
shown in Fig.~\ref{rescaledfs}.
\begin{figure}[htbp]
  \epsfxsize=14cm \hfil\epsfbox{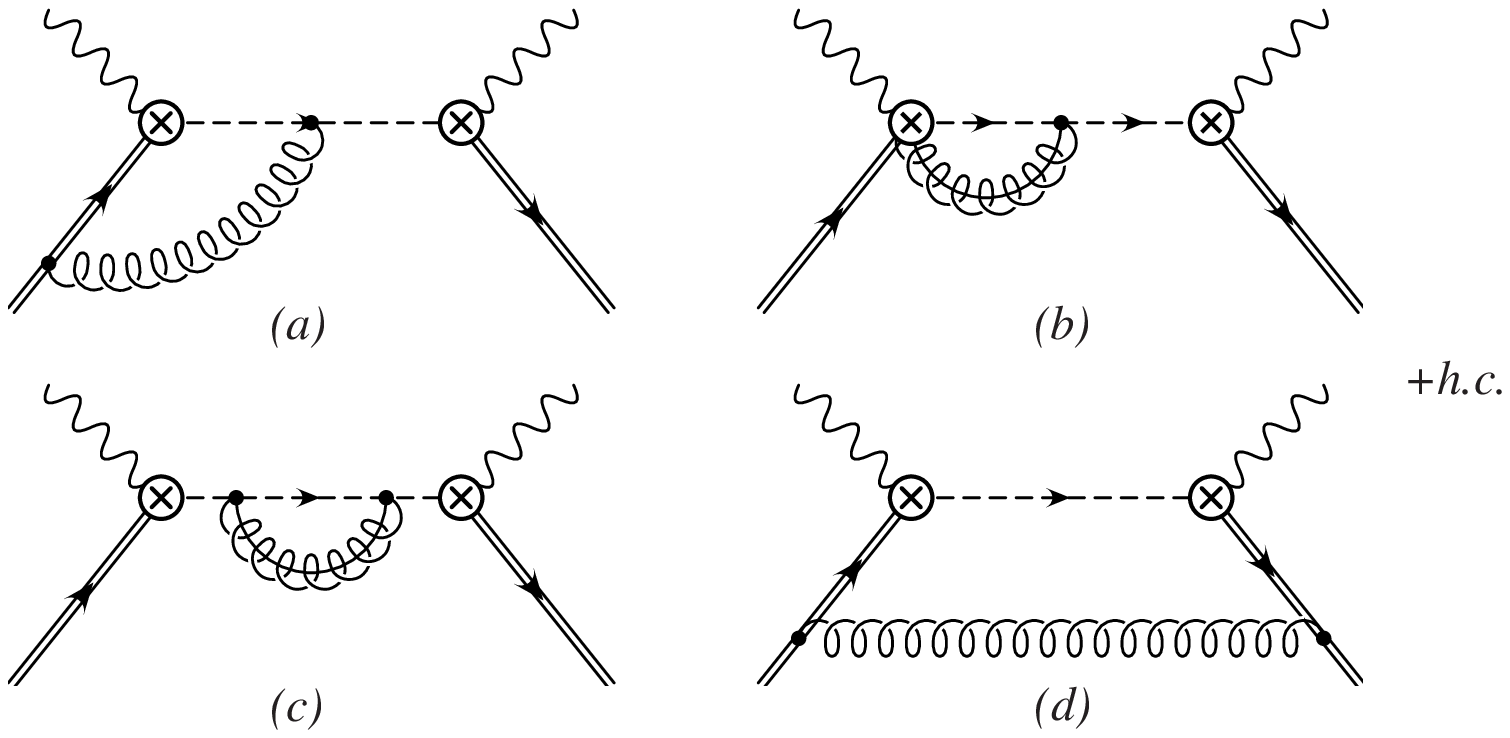}\hfill
     \caption{Collinear--soft theory Feynman diagrams which contribute
       to the forward scattering amplitude through $O(\alpha_s)$.}
     \label{rescaledfs}
\end{figure}
As with LEET, all divergences are regulated in dimensional
regularization.  Expanding the expression for the forward scattering
amplitude obtained from these graphs in powers of $(1-x+\hat k_+)$, we
find for the differential decay rate
\begin{eqnarray}
\label{diff_LCET}
\frac{d\Gamma}{dx}\Bigg|_{k_+} &=& C_V^2(\tilde\mu) \Gamma_0 \Bigg\{ \left[ 1 +
  \frac{\alpha_s C_F}{4 \pi} \left(\log^2 \frac{\tilde\mu^2}{m_b^2} + 5 \log 
  \frac{\tilde\mu^2}{m_b^2}  \finite{+ \frac{\pi^2}{6} + 7} \right)\right]
\delta(1-x+\hat k_+) \nonumber\\ 
&&\qquad
-  \frac{\alpha_s C_F}{4 \pi}\,\theta(1-x+\hat k_+)  \left[ 4
  \left(\frac{\log(1-x+\hat k_+)} 
{1-x+\hat k_+}
\right)_+ + 7
  \left(\frac{1}{1-x+\hat k_+}\right)_+ \right] \Bigg\}\,.
\end{eqnarray}
Comparing Eqs.\ (\ref{diff_LCET}) and (\ref{leetrate}) gives the
short-distance coefficient $C(y,x;\mu)$.  At tree level, the matching
is trivial, and we write
\begin{equation}
C(y,x;\tilde\mu)=C_V^2(\tilde\mu)\left[\delta(y-x)+{\alpha_s C_F\over 4\pi}
C^{(1)}(y,x;\tilde\mu)\right]+O(\alpha_s^2)\,,
\end{equation}
where $\mu$ is the matching scale.  At one loop, we find
\begin{eqnarray}
C^{(1)}(y,x;\tilde\mu)&=&\left( 2 \log^2 {\tilde\mu^2 \over m_b^2} + 
3 \log {\tilde\mu^2 \over m_b^2} \finite{+ \frac{\pi^2}{4} + 7 }
  \right) \delta(y-x) - \left( 3 + 4 \log {\tilde\mu^2 \over m_b^2} \right)
 \frac{\theta(y-x)}{(y-x)_+}
\nonumber\\
&&   + 4 \, \theta(y-x)
\left(\frac{\log(y-x)}{y-x} \right)_+\nonumber\\
&=&
\left(2\log^2{\tilde\mu^2\over m_b^2(y-x)}+3 \log{\tilde\mu^2\over m_b^2
(y-x)} \finite{+ \frac{\pi^2}{4} + 7} \right)\delta(y-x)\nonumber\\
&&
-4\,{\theta(y-x) \over y-x}\log{\tilde\mu^2\over m_b^2(y-x)}
-3\,{\theta(y-x) \over y-x}\,.
\label{LEETmatch}
\end{eqnarray}
At the scale $\tilde\mu = m_b \sqrt{y-x}$ the logarithmic terms
vanish, and we find
\begin{equation}
C^{(1)}(y,x;m_b\sqrt{y-x})= -3\,{\theta(y-x) \over y-x} \finite{+\left(
\frac{\pi^2}{4} + 7 \right) \delta(y-x)}\,.
\end{equation}
The matching scale is therefore different for each operator $O(y)$.

\OMIT{ Two features need to be pointed out. First, the matching scale
  depends on $y$, the light cone component of the residual momentum of
  the heavy quark inside the $B$ meson. Since we take $y$ to be a
  continuous index we have an infinite set of operators $O(y)$, where
  each operator in the set gets matched onto at a different scale.
  This is quite different from more standard effective theories and
  the implications need to be investigated in more detail.  Second,
  the dependence of the matching coefficient on $1/(y-x)$ implies that
  the single plus distributions can not be reproduced correctly in a
  resummed expression, without explicitly retaining such terms
  proportional to $1/(y-x)$. In this work we are only comparing to
  known results of large moments of the photon energy spectrum. As
  will be shown in the next section, this problem does not occur in
  this case.  The correct procedure to sum the plus functions directly
  will be explained elsewhere.}

\subsection{Renormalization group}

The differential decay rate in LEET given in (\ref{leetrate}) may be
written as
\begin{eqnarray}
\frac{d\Gamma}{dx}
&=&\Gamma_0 \int \!dy \, C(y,x;\tilde\mu) \left\{\left[1-{\alpha_s C_F\over 4
\pi}\left(
\log^2{\tilde\mu^2\over m_b^2(1-y+\hat k_+)^2}-2\log{\tilde\mu^2\over m_b^2
(1-y+\hat k_+)^2}\right)
\right]
\right.\nonumber\\
&&\times\delta(1-y+\hat k_+)+\left.{\alpha_s C_F\over
    4\pi}\left({4\over 1-y+\hat k_+}\log{\tilde \mu^2\over 
m_b^2(1-y+\hat k_+)^2}-{4\over 1-y+\hat k_+}\right)\right\}\,,
\end{eqnarray}
and so the large logarithms in the matrix element of $O(y; \tilde\mu)$ vanish
at the scale $\tilde\mu = m_b (1-y+\hat k_+)$.  (This expression looks
highly singular, but as can be seen from (\ref{leetrate}), the delta
functions combine with the other terms to form plus functions.)  Thus,
in order to sum all logarithms of $\mu$ we must continue to run the operator
$O(y)$ in LEET.
{}From (\ref{Orenormalization}) and
(\ref{softct}) we obtain the renormalization group equation
\begin{equation}
\mu \frac{d}{d\mu} C(y,x;\tilde\mu) = \int \! dy' \, \gamma(y,y';\tilde\mu)
C(y',x;\tilde\mu) \,,
\label{leetrge}
\end{equation}
where $\gamma(y,y^\prime;\tilde\mu)$ is the continuous anomalous dimension
matrix
\begin{equation}
\gamma(y,y';\tilde\mu) = {\alpha_s(\tilde\mu) C_F \over \pi}
\left[\left( \log{{\tilde\mu^2 
\over m^2_b}}-1\right)\delta(y'-y) - \frac{2}{(y'-y)_+} \theta(y'-y)
\right] \,.
\label{softanomalous}
\end{equation}
Solving (\ref{leetrge}) analytically, however, is nontrivial and
beyond the scope of this work \cite{llr}. Instead, we may
diagonalize the anomalous dimension matrix by taking high moments of
the spectrum.  This will allow us to compare our results to those of
Refs.~\cite{ks,ar}. Note that in Refs.~\cite{ks,ar} both leading and
next-to-leading logarithms were resummed. This requires the two loop
contribution to the $1/\epsilon^2$ counterterm, the full one loop matching
condition, and the two loop running of $\alpha_s$, none of which have
been included here. As a result our calculation only resums
the leading logarithms and a class of the subleading
logarithms. However, it is straightforward to extract from the
literature a resummation of exactly the same set of logarithms.

To calculate the moments we set the residual momentum $k$ to zero.
(This residual momentum can easily be incorporated by boosting from
the rest frame of the $b$ quark, $p_b = m_b v$, to the frame $p = m_b
v + k$). Taking moments unconvolutes the expression for the
differential decay rate in LEET (\ref{diff_convolution}) and we obtain
\begin{eqnarray}
\Gamma(N) &=& \int _0^1 dx\,x^{N-1} \frac{d \Gamma}{dx}\nonumber\\
&=& \Gamma_0\int _0^1 dx\,x^{N-1} \int_{-\infty}^\infty dy \, 
C(y-x;\mu) \langle
O(y;\mu) \rangle \nonumber\\
&=& \Gamma_0\int_0^1 dz \,z^{N-1} C^\prime (1-z;\tilde\mu) \int _0^1 dy
\,y^{N-1} \langle  O(y;\tilde\mu) \rangle \nonumber\\
&\equiv& \Gamma_0 \, C(N;\tilde\mu) \,\langle O(N;\tilde\mu) \rangle \,,
\label{momofdecayrate}
\end{eqnarray}
where we have used 
\begin{equation}
C(y-x) = \frac{1}{y}\, C^\prime \left(1-\frac{x}{y}\right)
\Theta(y-x)\,
\end{equation}
since $C(y-x)$ just contains delta functions and plus distributions.
Thus, the operator $O(N;\mu)$ is just a linear combination of the set
of operators $O(y;\mu)$. The matching from the collinear-soft theory
onto LEET at tree level is trivial, and we find
\begin{equation}
C(N;\tilde\mu) = C_V^2(\tilde\mu)\left[1  + \frac{\alpha_s C_F}{4 \pi}
C^{(1)}(N;\tilde\mu)\right] + O(\alpha_s^2)\,.
\end{equation}
Determining $C^{(1)}(N;\mu_0)$ requires the one-loop expression of
$\Gamma(N)$ in the collinear-soft theory and the one-loop matrix
element of $O(N;\mu)$ between partonic states.  The one-loop
expression for the differential decay rate in the collinear-soft
theory is given in (\ref{diff_LCET}). Setting $k_+$ to zero and taking
moments we obtain
\begin{eqnarray}\label{LCET_moment}
\Gamma(N) &=& \int_0^1 x^{N-1} \frac{d\Gamma}{dx}\nonumber\\
&=& \Gamma^0 \, C_V^2(\tilde\mu) \left\{ 1 - \frac{\alpha_s C_F}{4 \pi}
     \left[ 2 \log^2{N\over n_0} - 7 \log{N \over n_0} - \log^2 {\tilde\mu^2
         \over m_b^2} - 5 \log {\tilde\mu^2\over m_b^2}\right]
   \right\} + \ldots\,,
\end{eqnarray}
where $n_0 = e^{- \gamma_E}$. This needs to be compared to the one-loop
matrix element of $\langle O(N;\mu) \rangle$, which can be obtained
from (\ref{leetrate}):
\begin{equation}
\langle O(N;\tilde\mu) \rangle = \int^1_0 dy\, y^{N-1} \langle O(y;\tilde\mu) \rangle  = 
1 -
\frac{\alpha_s C_F}{4 
  \pi} \left[ 4 \log^2{\tilde\mu N\over 
     m_b n_0} - 4 \log{\tilde\mu N \over m_b n_0} \right] + \ldots\,.
\label{LEETN}
\end{equation}
The one loop matching coefficient is now easily determined using
(\ref{momofdecayrate}), (\ref{LCET_moment}) and (\ref{LEETN}) and we
find
\begin{equation}
C^{(1)}(N;\tilde\mu) =
   \frac{\alpha_s C_F}{4 \pi} \left[ 
   2\log^2 {\tilde\mu^2 N \over m_b^2 n_0} + 3 \log {\tilde\mu^2 N
     \over m_b^2 n_0} 
\right] + \ldots \,.
\end{equation}
At the matching scale $\tilde\mu=m_b\sqrt{n_0/N}$ all logarithms in
this matching coefficient vanish. Furthermore, from (\ref{LEETN}) it
is clear that the
matrix element $\langle O(N;\tilde\mu) \rangle$ contains no large logarithms
of $N$ at the scale $\tilde\mu = m_b n_0/N$. Thus we run in the
collinear-soft theory from $m_b$ to $m_b\sqrt{n_0/N}$, perform the
OPE, and run $C(N;\tilde\mu)$ from $m_b \sqrt{n_0/N}$ to $m_b n_0/N$.

The running of the coefficient $C_V$ in the collinear-soft theory from
the scale $m_b$ to the scale $m_b \sqrt{n_0/N}$ is obtained by setting
$\tilde\mu = m_b \sqrt{n_0/N}$ in (\ref{cvmu}). The running in LEET is
determined by the RGE for $C(N;\tilde\mu)$
\begin{equation}
\mu \frac{d}{d\mu} C(N;\tilde\mu ) = \gamma(N;\tilde\mu)\, C(N;\tilde\mu) \,,
\end{equation}
where the anomalous dimension is given by
\begin{equation}
\gamma(N;\tilde\mu) = \int^1_0 dz \; z^{N-1} \gamma (z;\tilde\mu) = 
- {\alpha_s(\tilde\mu) C_F \over \pi}\left[1- 2 \log{\left(\tilde\mu N \over
m_b n_0\right)}  \right] \,.
\end{equation}
The solution to this equation is
\begin{equation}
C(N;{m_b n_0 \over N}) = C^2_V \left( m_b \sqrt{{n_0 \over N}} \right) 
\left({\alpha_s(m_b {n_0 \over N}) \over \alpha_s(m_b \sqrt{{n_0 \over
N}})} \right)^{{2 C_F \over \beta_0}\left( 1 + {4 \pi \over \beta_0
\alpha_s} - 2 \log{N \over n_0}\right)} \left( {n_0 \over N} \right)^{2
C_F \over \beta_0} \,.
\end{equation}
This sums perturbative logarithms of $N$ into the coefficient $C(N)$.
We can then substitute this into (\ref{momofdecayrate}) to obtain an
expression for the resummed moments of the differential decay rate.

Using the result for $C_V(\mu)$ given in (\ref{cvmu}) and taking the
matrix element of $O(N;\mu)$ between hadronic states, we find find the
resummed expression for large photon energy moments of the decay $B
\to X_s \gamma$
\begin{equation}
\Gamma(N) = \Gamma_0 f\left(N;m_b n_0/N \right)
\left({\alpha_s(m_b \sqrt{{n_0 \over N}}) \over
\alpha_s}\right)^{{C_F \over \beta_0}\left( 5 - {8 \pi \over \beta_0
\alpha_s}\right)}
\left({\alpha_s(m_b {n_0 \over N}) \over \alpha_s(m_b \sqrt{{n_0 \over
N}})} \right)^{{2 C_F \over \beta_0}\left( 1 + {4 \pi \over \beta_0
\alpha_s} - 2 \log{N \over n_0}\right)} \,.
\label{Nsolution}
\end{equation}
Logarithms are explicitly summed in this expression and only long
distance physics is contained in the function $f(N;m_b n_0/N)$.

We can easily compare our results to those in the literature. A
resummed expression for $\Gamma(N)$ is given in Ref.~\cite{ks}:
\begin{equation}
\Gamma(N) = \Gamma_0 f(N;m_b/N) \;
{\rm exp} \left[ - \int^1_{n_0/N} {dy \over y} \left( 2
\int^{m_b\sqrt{y}}_{m_b y} {d \mu \over \mu} \Gamma_c(\mu ) +
\Gamma(m_b y) + \gamma(m_b \sqrt{y}) \right)\right] \,,
\label{ffrun}
\end{equation}
where, at one loop,
\begin{equation}
\Gamma_c(\mu) = {\alpha_s(\mu) C_F \over \pi}, \qquad  \Gamma(\mu) =
-{\alpha_s(\mu) C_F \over \pi}, \qquad \gamma(\mu) = - {3
\alpha_s(\mu) C_F \over 4 \pi} \,.
\end{equation}
Note that the cusp anomalous dimension $\Gamma_c(\mu)$ is the
contribution to the anomalous dimension from the $1/\epsilon^2$
counterterm. Using only the one loop cusp anomalous dimension,
tree level matching, and the one loop running of $\alpha_s$, 
Eq.~(\ref{ffrun}) resums leading logarithms and the same class of
next-to-leading logarithms we resum in our calculation. Performing the
integrals in the exponent we reproduce (\ref{Nsolution}). 
Thus the approach presented here, based on an effective field theory,
is in agreement with the factorization formalism approach for summing
perturbative logarithms. 

\section{Conclusions}

{}In the specific case of $\bar B \to X_s \gamma$ we have shown how
Sudakov logarithms can be summed within an effective field theory
framework.  First we construct an intermediate theory which includes
both collinear and soft degrees of freedom. By performing a one-loop
calculation we show that this collinear-soft theory can be matched
onto QCD at the scale $m_b$ without introducing logarithmic terms into
the short-distance coefficient. In addition we determine the one-loop
anomalous dimension and solve the RGE. Next we integrate out collinear
modes at the scale $m_b \sqrt{y-x}$ by switching to LEET. We perform
an OPE in powers of $(y-x)$ which leads to the appearance of a
nonlocal operator where two vertices are separated along the
light-cone. The matrix element of this operator between $B$ meson
states is the structure function. We perform the OPE at one-loop in
the collinear-soft theory and match onto the nonlocal operator in
LEET. At the scale $m_b \sqrt{y-x}$ no logarithmic terms are
introduced into the short-distance coefficient.

In order to compare to the factorization formalism results in the
literature we repeat our analysis for large moments of the decay rate.
In this case we find that the collinear-soft theory matches onto LEET
at the scale $m_b\sqrt{n_0/N}$, and that there are no large logarithms
in the matrix element of the bilocal operator at the scale $m_b
n_0/N$. Using the renormalization group equations in the
collinear-soft theory we sum logarithms of $N$ between the scales
$m_b$ and $m_b\sqrt{n_0/N}$. We then switch to LEET and sum logarithms
of $N$ between the scales $m_b\sqrt{n_0/N}$ and $m_b n_0/N$. This sums
all perturbative logarithms of $N$. We find that our result agrees
with results presented in the literature. This gives us confidence
that we have constructed the correct effective field theory.

Though we have presented this work entirely in the context of $B \to
X_s \gamma$ our approach is general. It should be straightforward to
apply the collinear-soft theory and LEET to other processes in which 
Sudakov logarithms
arise.  Furthermore, this approach could also be applied to exclusive
decays, in which case one does not perform the final OPE onto LEET,
but remains in the collinear-soft theory.  This could be applied to
recent results on factorization in nonleptonic decays
\cite{bbns}, as well as 
LEET-based relations between form-factors in decays to highly energetic
final states\cite{cyopr}.  Since these latter results depend only on the
spin symmetry of LEET, which is also present in the collinear-soft 
theory, they should remain valid in the present approach.

\section*{Acknowledgements}

We thank Craig Burrell, Adam Leibovich, Aneesh Manohar, Tom Mehen, Dan
Pirjol, Ira Rothstein and Iain Stewart for useful discussions.  This 
work was supported in part by the Natural Sciences and Engineering Research
Council of Canada and the Sloan Foundation.


\end{document}